\def\year{2022}\relax
\documentclass[letterpaper]{article} % DO NOT CHANGE THIS
\usepackage{aaai22}  % DO NOT CHANGE THIS
\usepackage{times}  % DO NOT CHANGE THIS
\usepackage{helvet}  % DO NOT CHANGE THIS
\usepackage{courier}  % DO NOT CHANGE THIS
\usepackage[hyphens]{url}  % DO NOT CHANGE THIS
\usepackage{graphicx} % DO NOT CHANGE THIS
\usepackage{natbib}  % DO NOT CHANGE THIS AND DO NOT ADD ANY OPTIONS TO IT
\usepackage{caption} % DO NOT CHANGE THIS AND DO NOT ADD ANY OPTIONS TO IT
\DeclareCaptionStyle{ruled}{labelfont=normalfont,labelsep=colon,strut=off} % DO NOT CHANGE THIS
\usepackage{algorithm}
\usepackage{algorithmic}

\usepackage{graphicx}
\usepackage{amsmath}
\usepackage{amssymb}
\usepackage{subfig}
\usepackage{cleveref}

\usepackage{amsfonts}
\usepackage{mathrsfs}
\usepackage{dsfont}

\usepackage{enumitem}

\usepackage{amsmath}

\usepackage{newfloat}
\usepackage{listings}
\floatstyle{ruled}
\newfloat{listing}{tb}{lst}{}
\floatname{listing}{Listing}
\title{Statistics of extreme events in coarse-scale climate simulations via machine learning correction operators trained on nudged datasets}
\author{A. Charalampopoulos$^{1}$\footnote{Corresponding authors: alexchar@mit.edu, sapsis@mit.edu}, S. Zhang$^{2}$, B. Harrop$^{2}$, L. R. Leung$^{2}$, and T. P. Sapsis$^{1*}$\\
}
\affiliations{
    \textsuperscript{\rm 1}Department of Mechanical Engineering, Massachusetts Institute of Technology\\
    Cambridge, MA 02139, USA\\
    \textsuperscript{\rm 2}Pacific Northwest National Laboratory, Richland, WA 99354, USA\\    
}

\usepackage{bibentry}
\begin{document}

\maketitle

\begin{abstract}
This work presents a systematic framework for improving the predictions of statistical quantities for turbulent systems, with a focus on correcting climate simulations obtained by coarse-scale models. Specifically, failure to incorporate all relevant scales in climate simulations leads to discrepancies in the energy spectrum as well as higher order statistics. While high resolution simulations or reanalysis data are available, at least for short periods, they cannot be directly used as training datasets to machine learn a correction for the coarse-scale climate model outputs, since chaotic divergence, inherent in the climate dynamics, makes datasets from different resolutions incompatible. To overcome this fundamental limitation we employ coarse-resolution model (here we employ Energy Exascale Earth System Model, E3SM) simulations nudged towards high quality climate realizations, here in the form of ERA5 reanalysis data. The nudging term is sufficiently small to not “pollute” the coarse-scale dynamics over short time scales, but also sufficiently large to keep the coarse-scale simulations “close” to the ERA5 trajectory over larger time scales. The result is a “compatible" pair of the ERA5 trajectory (used as output training data) and the weakly nudged coarse-resolution E3SM output that is used as input training data to machine learn a correction operator. We emphasize that the nudging step is used only for the training phase. Once training is complete, we perform free-running coarse-scale E3SM simulations without nudging and use those as input to the machine-learned correction operator to obtain high-quality (corrected) outputs. The model is applied to atmospheric climate data with the purpose of predicting global and local statistics of various quantities of a time-period of a decade. Using ERA5 datasets that are not employed for training, we demonstrate that the produced datasets from the ML-corrected coarse E3SM model have statistical properties that closely resemble the observations. In particular, the corrected coarse-scale E3SM output closely captures the non-Gaussian statistics of quantities such as temperature, wind speed and humidity,  as well as the frequency of occurrence of extreme events, such as tropical cyclones and atmospheric rivers. We present thorough comparisons and discuss limitations of the approach. 
\end{abstract}

\section{Problem Formulation}
Accurate statistical climate predictions require high-fidelity simulations that come with large computational cost. As a result, improving upon the predictions of coarse-scale climate models has become a critical goal in order to develop credible climate scenarios. The developed method aims to augment the accuracy of statistical properties of coarse-scale free-running (i.e. without the influence of observations which are obviously not available for future projections) climate models, via corrections based on past data. Despite recent successes in using online correction terms in the evolution equations of the model~\cite{yuval2020stable, yuval2021use, sanford2022improving, charalampopoulos2022uncertainty}, many such implementations face severe stability issues. To circumvent this obstacle, a non-intrusive approach is developed. Hence, after a free-running coarsely resolved climate output has been generated, the hybrid approach corrects the model output in a post-processing manner. A reference dataset is selected for testing the effectiveness of the scheme, the ERA5 dataset~\cite{hersbach2020era5}. 
\begin{figure*}
    \centering
    {{\includegraphics[width=0.7\textwidth]{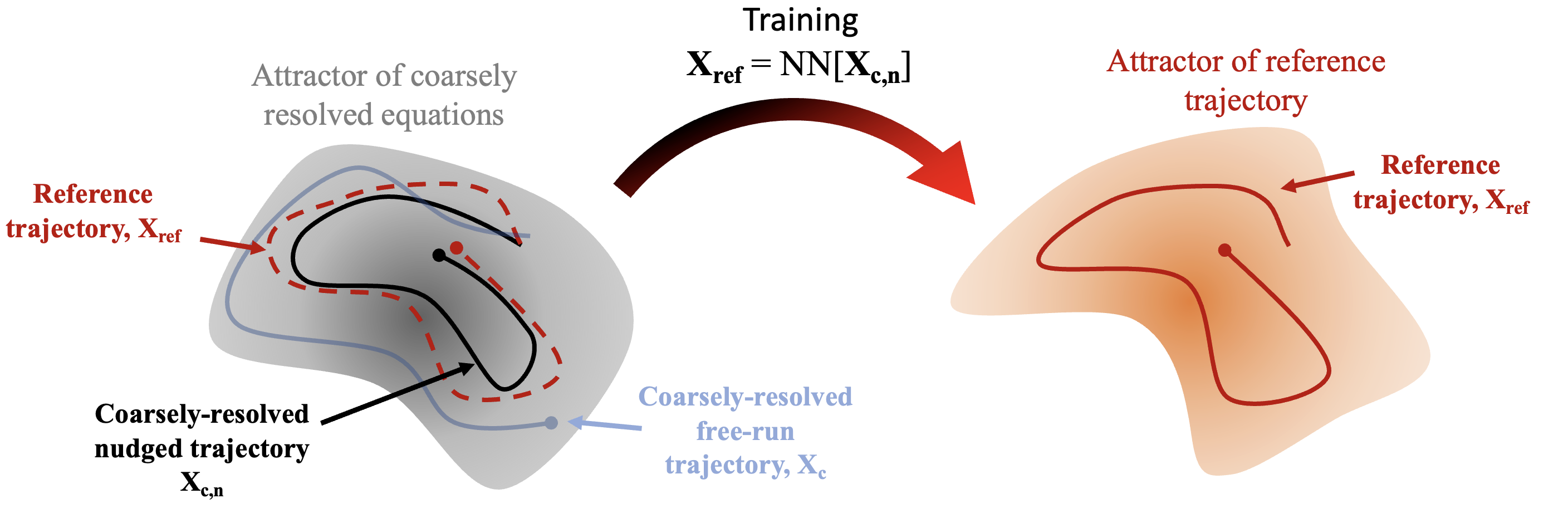} }}%
    \caption{\textbf{Description of the method } that learns a map between the attractor of the coarsely-resolved equations and the attractor of the reference trajectory. Left: the red dashed curve represents a reference trajectory (here ERA5). The black curve is a coarsely-resolved nudged trajectory towards the reference trajectory. The blue curve is the free-run coarsely-resolved trajectory that is not used for training (shown for reference). Right: the target attractor and the target trajectory (red), same as the dashed curve shown at the left plot. For training we use the coarsely-resolved nudged trajectory as input and the reference trajectory as output to machine learn a map between them. After we obtain the map we use as input coarsely-resolved free-run simulations (blue) and obtain a trajectory that accurately captures the shape of the target attractor.}%
    \label{fig:Attractor_Plot}%
\end{figure*}
For the generation of coarse-scale climate data, the atmospheric component of the Energy Exascale Earth System Model (E3SM) is used. In particular, version 2 of the E3SM Atmospheric Model (EAMv2)~\cite{dennis2012cam,taylor2009non, golaz2022doe}. Appropriate boundary conditions over the Earth's surface are prescribed~\cite{oleson2013technical}. Simulations are run on an unstructured grid of approximately $1^{\text{o}}$($\sim 110[\text{km}])$ resolution per sigma-level and 72 levels along the vertical direction. The vertical levels extend from the Earth surface up to altitude of about $64 [\text{km}]$, corresponding to $\sim 0.1 [\text{hPa}]$. The evolution model equations for the coarse-scale model have the form
\begin{gather}
\begin{split}
    \frac{\partial \mathbf{X}_c}{\partial t} = \mathcal{D}\left( \mathbf{X}_c \right) + \mathcal{P} \left(\mathbf{X}_c \right),
\end{split}
\end{gather}
where $\mathbf{X}_c = \left(U, V, T, Q\right)$ represent the set of coarsely-resolved system variables, $\mathcal{D}$ is the operator containing the dynamics of the system~\cite{zhang1995sensitivity,golaz2002pdf} and $\mathcal{P}$ is the operator concerning the physics of the system~\cite{morrison2008new,liu2016description,mlawer1997radiative}. Variables $(U,V)$ correspond the zonal and meridional components of wind velocity, $T$ is wind temperature and $Q$ is specific humidity. From here on now, the coarse-scale free-running dataset will be labeled as CLIM and will be denoted as $\mathbf{X}_c$. 

For reference data, denoted as $\mathbf{X}^{\text{ref}}$, ERA5 reanalysis data is used, which is projected onto the coarse unstructured grid of EAMv2. The datasets discussed herein contain information from 2007-2017. For this timescale the studied climate systems can be assumed to be in a statistical steady state.

Since the goal of the approach is to correct the long-time statistics of coarse-scale climate simulations in a post-processing manner, it is important to isolate the main discrepancies between the coarse-scale simulations and the reference data that are responsible for these differences. In general, discrepancies between two turbulent simulations, one high-fidelity (i.e. reference) and one free-running coarse, can be grouped into two categories: (i) discrepancies due to chaotic divergence; (ii) discrepancies due to deformation of the attractor due to coarse-scale resolution. 
\par
Chaotic divergence is an intrinsic property of turbulent systems. It can be observed even between two solutions of the same dynamical system, with ever slightly different initial conditions. It is a manifestation of the fact that by definition, at least one of the eigenvalues of the linear part of the system is positive. As a result, infinitesimal energy transferred to perturbations along these directions will result in finite magnitude perturbations. The system is allowed to equilibrate with the intervention of nonlinear terms that will transfer this energy from the unstable perturbations to stable ones. However, the two deviating trajectories will still remain on the same attractor and thus retain the same statistical properties. Therefore these chaos-induced discrepancies should not contribute to the correction scheme for long-time statistics.
\par
On the other hand, difference in long-time statistics implies a different statistical steady-state and thus different attractors. These intrinsic dynamical differences between the simulations produce energy discrepancies in various scales between the produced datasets. It is exactly these corrections we aim to learn and fix. 

\begin{figure*}
    \centering
    {{\includegraphics[width=0.7\textwidth]{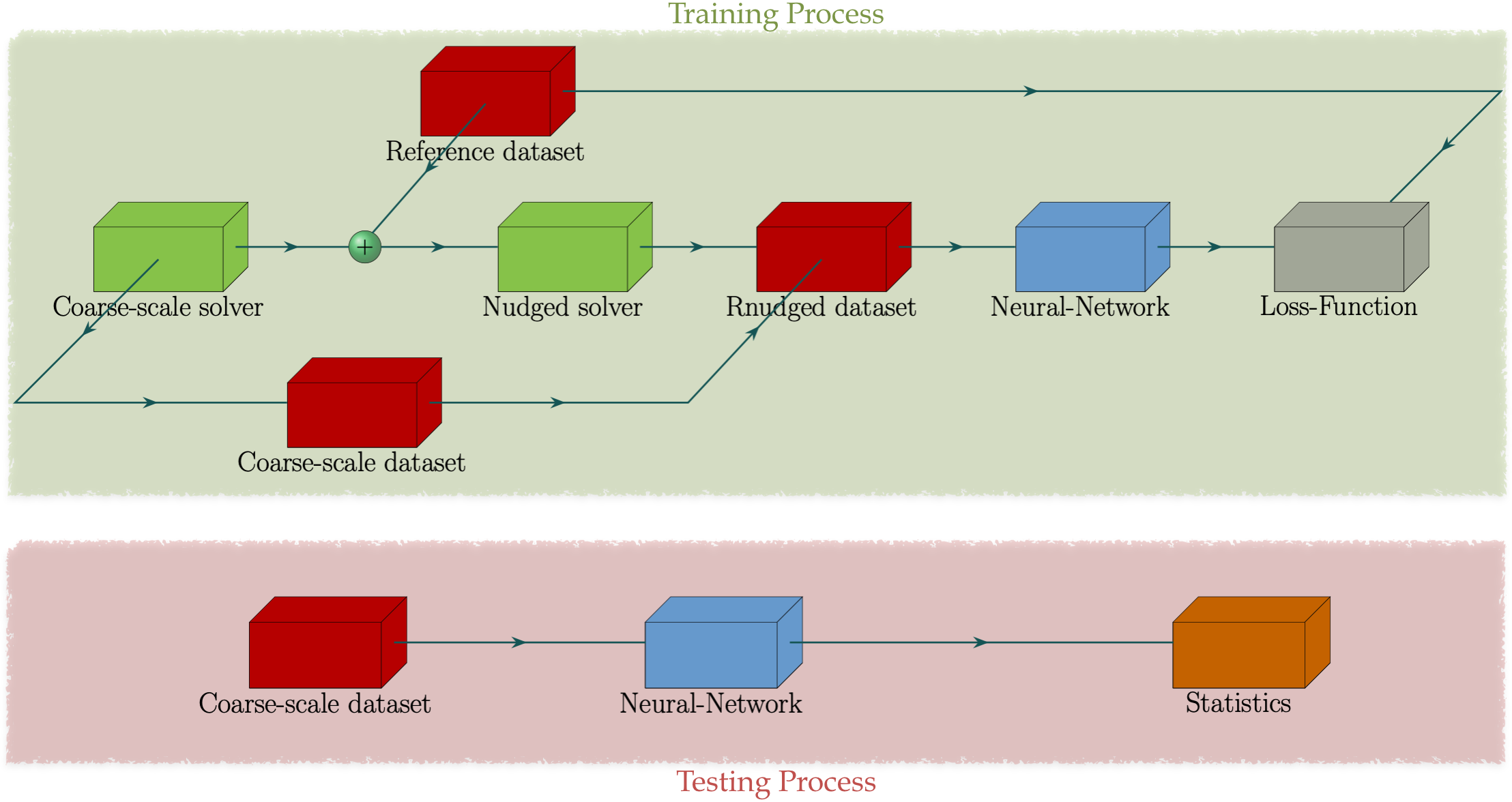} }}%
    \caption{Schematics of the training process (top) and testing process (bottom), for the non-intrusive hybrid method.}%
    \label{fig:LSTM_Generic_Methodology}%
\end{figure*}

Given the two previous observations, it is clear that it is not possible to use a dataset of free-running climate simulation (CLIM) and one of ERA5 and try to machine learn a map between the two, i.e. a map that takes as input a CLIM timeseries and produces as output an ERA5 timeseries. To eliminate the problematic component, i.e. chaos-induced divergence we design a new CLIM dataset (we call it nudged CLIM and denote it as $\mathbf{X}_{c,n}$). Ideally, one can produce a dataset that is preserving the coarse-scale behavior of the climate model but does not suffer from the chaos-induced divergence with the ERA5. To this end, the concept of nudging, that has been used extensively in the context of data assimilation~\cite{sun2019impact,zhang2022further}, is employed.  Specifically, we utilize the EAMv2 solver (generator of the CLIM dataset) with an extra term, the nudging term, that is `pulling' the CLIM solution close to the ERA5 solution:
\begin{gather}
\begin{split}
    \frac{\partial \mathbf{X}_{c,n}}{\partial t} = \mathcal{D}\left( \mathbf{X}_{c,n} \right) + \mathcal{P} \left(\mathbf{X}_{c,n} \right) +\mathcal{N} \left( \mathbf{X}_{c,n}; \mathbf{X}^{\text{ref}}\right) ,
\end{split}
\end{gather}
where the relaxation term $\mathcal{N}$ is called the nudging tendency and it corrects the coarse-scale solution based on the ERA5 reference solution. In this study, the nudging tendency $\mathcal{N}$ is given by the algebraic term
\begin{gather}
\begin{split}
    \mathcal{N}\left( \mathbf{X}_{c,n}-\mathbf{X}^{\text{ref}} \right) = -\frac{1}{\tau} \left( \mathbf{X}_{c,n} -\mathcal{H} \left[ \mathbf{X}^{\text{ref}} \right] \right).
\end{split}
\end{gather}
Parameter $\tau$ is a relaxation timescale that has a large value (so that $1/\tau$ is small compared with the other terms in the equation), and $\mathcal{H}$ is an operator that maps $\mathbf{X}^{\text{ref}}$ to the coarse resolution. A schematic of the proposed mapping learned during training can be seen in~\cref{fig:Attractor_Plot}.

The resulted nudged trajectory (black curve on the right panel) is subjected to this very small perturbation, the nudging term that is keeping it close to the reference trajectory, i.e. the ERA5 trajectory (red dashed line). Moreover, because the overall magnitude of the nudging term is very small the long-time statistics of the coarsely resolved before-nudge trajectory should be close to that of the free-running coarsely-resolved trajectory that starts from the before-nudge state (shown with blue color). The latter will naturally diverge from the ERA5 if not continuously being nudged due to chaotic properties, even if it was initiated very close to ERA5. Having the before-nudge trajectory we can now use it to machine learn a scheme that will map it to the reference trajectory, i.e.the ERA5. That is the basic approach of our framework. We emphasize that nudging is used ONLY for the generation of training data. Once the map has been trained we will feed it with free-run CLIM simulations (i.e. free-run coarse-scale climate simulations without nudging) to obtain outputs that have corrected long-time statistics, i.e. represent the target attractor accurately.

The resulting training and testing process are described in~\cref{fig:LSTM_Generic_Methodology}. During training, the Nudged EAMv2 solver is used to produce training data. After a spectral correction of the data (described in the next section), the resulting R-nudged dataset is used as input for the neural network. The neural network then learns a mapping between the reference ERA5 data and the input R-nudged dataset. The neural network used is described in detail in subsection `Neural Network Architecture'. During testing, EAMv2 is used to generate a free-running coarse-scale dataset. This dataset is used as input to the trained neural network which produces a corrected dataset with the desired statistics. \textit{Hence, during testing, the model is not assessed in its ability to mimic the reference data snapshot-by-snapshot but by its ability to learn its underlying statistics}. 

Revisiting the nudging procedure, parameter $\tau$ is chosen so the nudged solution $\mathbf{X}_c$ satisfies two properties: (a) it reduces the divergence of the nudged simulation from the reference solution (i.e., ERA5) $\mathcal{H} \left[ \mathbf{X}^{\text{ref}} \right]$, i.e. allowing for a generalizable mapping between the two datasets; (b) it resembles the statistical properties of the coarse-scale free-running simulation. The second property is important in the context of machine learning to ensure that the learned mapping during training will be applicable while testing using CLIM data. This implies that the attractor of the Nudged simulation has the same shape as the attractor of CLIM. However, no parameter $\tau$ can be found that explicitly satisfies this condition. This is due to the arbitrariness of the algebraic form of the nudging term. While an algebraic term is easily implemented it yields a constant dissipation rate across all wavenumbers which are in general not consistent with the dynamics of the system. This leads to suppression of extreme events and thus statistics with less heavy tails. A remedy for this issue is shown in subsection `Spectral Correction of Nudged Dataset', where the energy spectrum of nudged simulations is brought closer to that of CLIM. One other possible remedy is changing the form of the nudging term via an appropriate energy balance argument.

\subsection{Spectral Correction of Nudged Dataset}\label{subsec:Rnudging}
\begin{figure*}
    \centering
    {\includegraphics[width=0.90\textwidth]{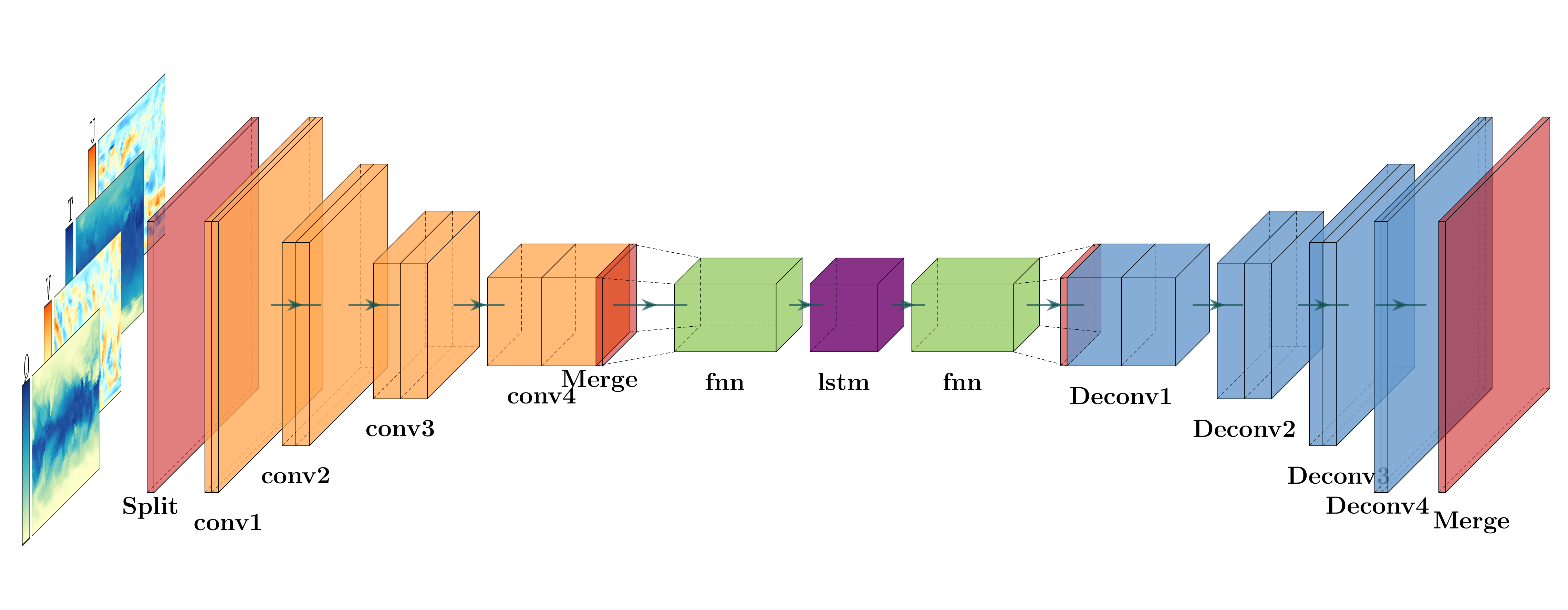} }%
    \caption{Neural network architecture of the non-instrusive model for a training on a particular sigma-level.}%
    \label{fig:LSTM_Architecture}%
\end{figure*}
As described in the previous subsection, the nudged dataset $\mathbf{X}_{c,n}$ is used during training to eliminate discrepancies due to chaotic divergence between input data and ERA5 reference data. However, for the learned mapping to be completely transferable to the free-running $\mathbf{X}_c$ testing data, the two datasets must have  attractors that are as similar as possible. If not, these discrepancies will manifest themselves to the target of this work, i.e. tails of statistics of various quantities. This is a result of discrepancies in the energy spectrum of the nudged solution with respect to the coarse-scale solution. These energy spectra differences lead to different statistical steady-state behaviours of testing data $\mathbf{X}_c$ and training data $\mathbf{X}_{c,n}$, which inhibits the generalizability of a model~\cite{shalev2014understanding}. 

%These discrepancies cannot be reconciled by simply choosing an appropriate $\tau$ as algebraic nudging adds linear dissipation to the system, thus always changing the energy spectrum of the resulting flow. 
%\par
To remedy the energy spectra differences between the testing input data $\mathbf{X}_c$ and training input data $\mathbf{X}_{c,n}$, a non-intrusive correction is proposed. The process is called `Reverse Spectral Nudging' with its purpose being to match the energy spectrum of the nudged solution to that of the coarse-scale solution. This modification implies that the statistics of the steady state of the two input datasets are very close to one another and thus generalizability improves. Hence, while traditional nudging schemes correct the coarse-scale solution with data from the reference solution, the proposed scheme further processes the nudged data by matching its energy spectrum to that of the corresponding free running coarse-scale flow. The corrected nudged data is termed as $\mathbf{X}_{c,rsn}$ and defined as
\begin{gather}\label{eq:R_nudge}
\begin{split}
    \mathbf{X}_{c,rsn}\left(\phi, \theta, t; k\right) = \sum_{m,n} R_{m,n} \{\hat{\mathbf{X}}_c(t)\}_{m,n} e^{i\left( m \phi +n \theta \right)}, ,
\end{split}
\end{gather}
where $\{\hat{\mathbf{X}}_c(t)\}_{m,n}$ are the spatial Fourier coefficients of $\mathbf{X}_{c,n}$, $\phi$ and $\theta$ corresponds to longitute and latitude respectively, while parameter $k$ denotes the number of the sigma-level. We also have   
\begin{gather}\label{eq:Rcoeff}
\begin{split}
    R_{k,l} = \sqrt{\frac{\mathcal{E}^{\text{coarse}}_{k,l}}{\mathcal{E}^{\text{nudge}}_{k,l}}}, 
\end{split}
\end{gather}
and
\begin{gather}\label{eq:Rcoeff2}
\begin{split}
    \mathcal{E}_{k,l} = \frac{1}{T} \int_0^T \hat{E}_{k,l}(t) \mathrm{d}t=\frac{1}{T} \int_0^T |\{\hat{\mathbf{X}}_c(t)\}_{k,l}|^2 \mathrm{d}t,
\end{split}
\end{gather}
where $T$ is the duration of the available data.

\subsection{Neural Network Architecture}\label{subsec:NN_Architecture}
In the current implementation, training is done on an level-by-level basis, where level here denotes atmospheric sigma-levels~\cite{taylor2020energy}. A schematic of the configuration for training on a particular layer is shown in~\cref{fig:LSTM_Architecture}. The model receives as input the predictive variables $\mathbf{X} = \mathbf{X}(\phi,\theta,t;k)$, where $\phi$ is the longitudinal angle and $\theta$ the latitudinal one. Snapshots of the entire horizontal discretization of the layer are used. Afterwards, a custom "split" layer separates the input into non-overlapping subregions. These subregions are periodically padded via a custom padding process, tasked with respecting the spherical periodicity of the domain. Then, each subregion is independently passed through a series of convolutional layers. The purpose of this process is to extract anisotropic local features in each subregion. 

%Each convolutional layer utilizes a rectified linear activation function (ReLU). This activation function allows for deriving features that are non-zero only when a particular criterion is met. This behaviour is vital for creating feature that pertain to extreme events like cyclones and atmospheric rivers.
%\par
Afterwards, the local information extracted from each subregion is concatenated in a single vector via a custom `merge' layer. The global information is now passed through a linear fully-connected layer, that acts as a basis projection of the spatial data onto a reduced-order latent space. The latent space data are then corrected by a long short-term memory (LSTM) layer~\cite{hochreiter1997long}. Subsequently they are projected back to physical space via another linear fully-connected layer. Afterwards, global information is split into the same subregions of the input, and distributed to a series of independent deconvolution layers that upscale the data to the original resolution. Finally, a custom `merge' layer gathers the information from each subregion and produces the final corrected snapshot.
\par
The motivation behind using LSTM neural networks lies in their ability to incorporate (non-Markovian) memory effects into the reduced-order model. This ability stems from Takens embedding theorem~\cite{takens1981detecting}. This theorem states that given delayed embeddings of a limited number of state variables, one can still obtain the attractor of the full system for the observed variables. This approach is known to be capable of improving predictions of reduced-order models~\cite{ vlachas2018data, charalampopoulos2022machine, wan2021}. 
\begin{figure*}
    \centering
    \subfloat[]{{\includegraphics[width=0.40\textwidth]{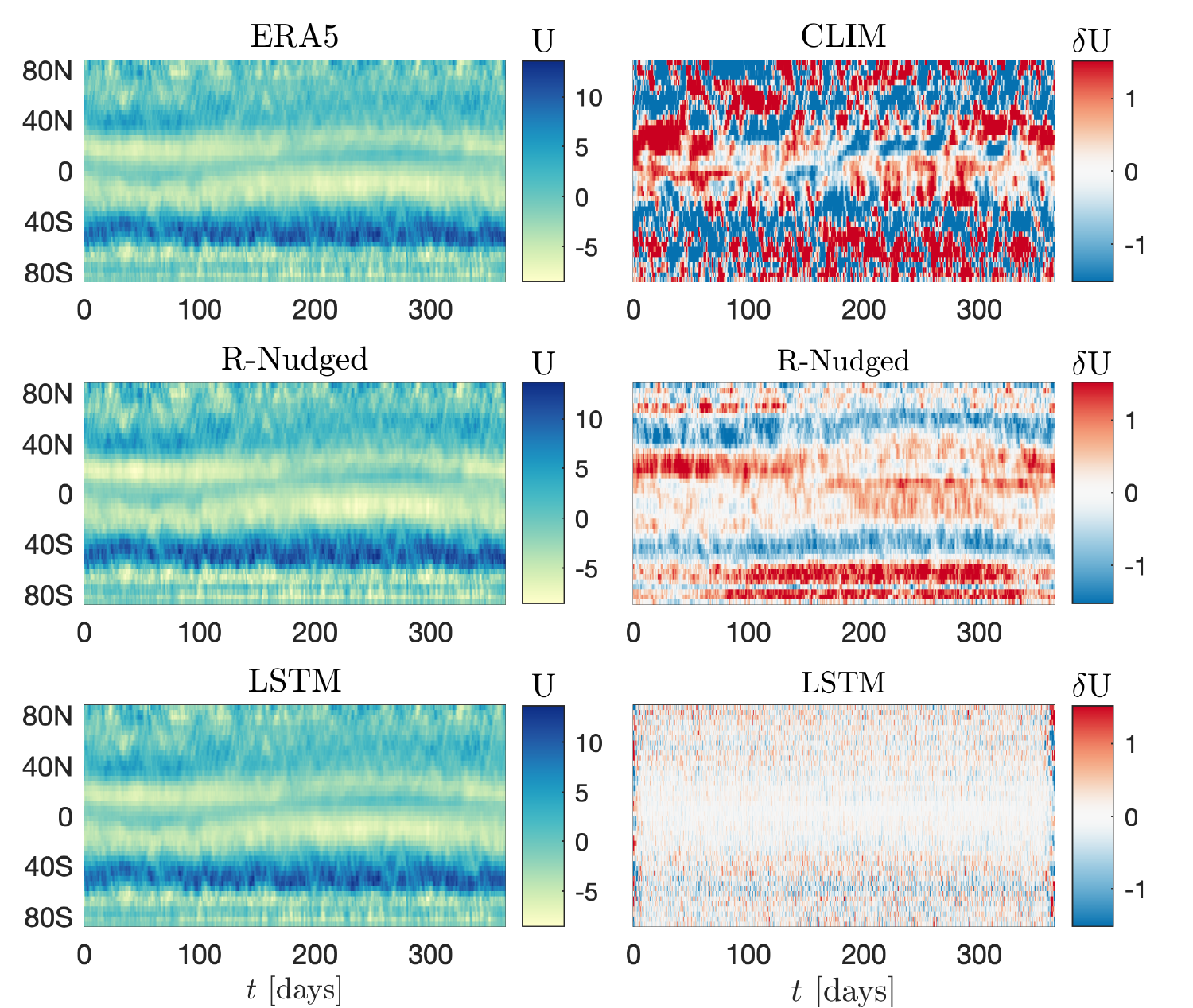} }}%
    %\qquad
    \subfloat[]{{\includegraphics[width=0.40\textwidth]{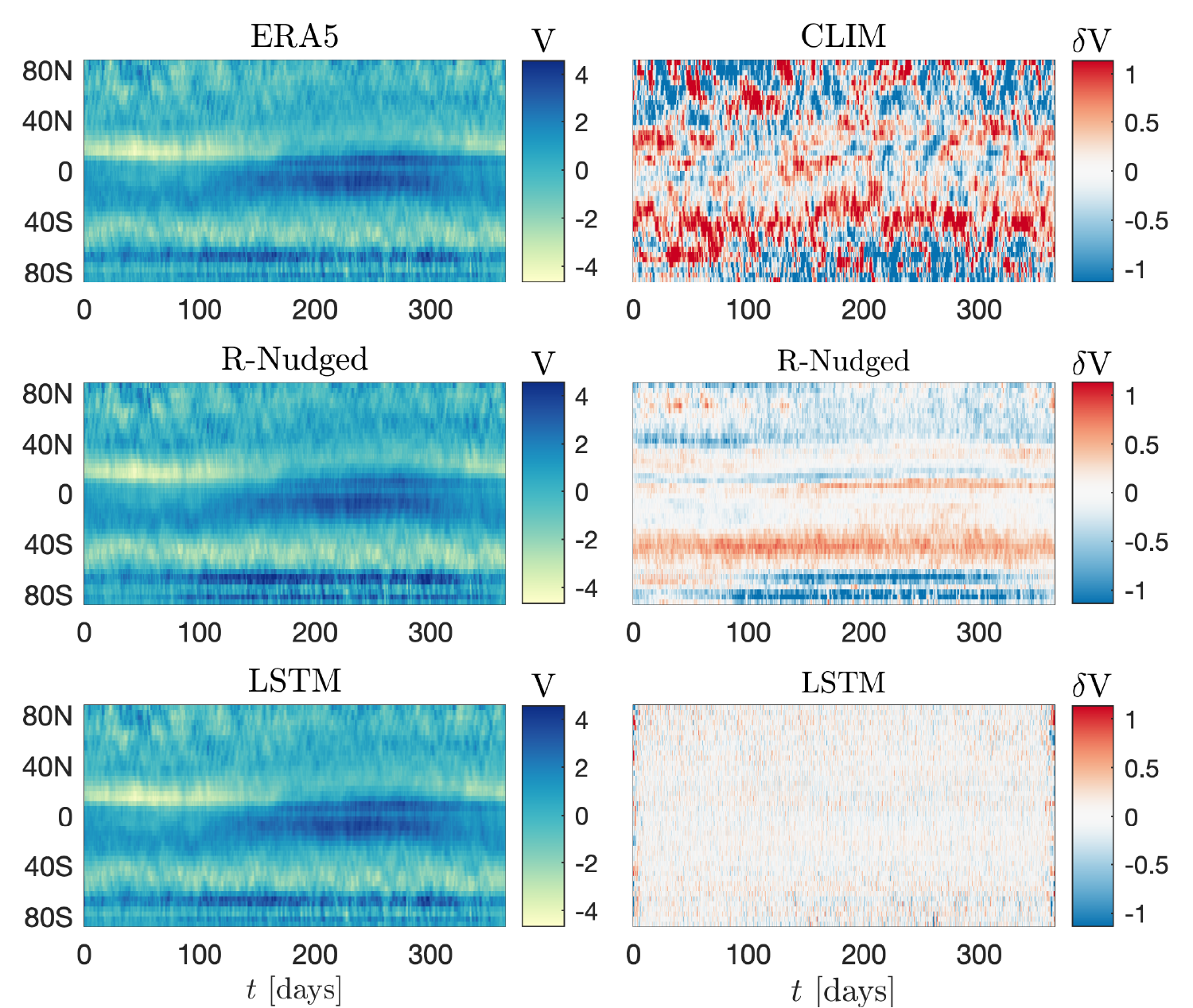} }}%
    \caption{(a) Zonally averaged predictions for $U$ (left column) and biases with respect to ERA5 predictions (right column). (b) Zonally averaged predictions for $V$ (left column) and biases with respect to ERA5 predictions (right column). Results are shown for near-surface data.}%
    \label{fig:UV_Zonal_Averages}%
\end{figure*}
In addition to temporal nonlocality, the model is nonlocal in space. Note, that in terms of the LSTM layer, this information comes in the form of the latent space coefficients, which in general correspond to global modes that correspond to rows of the fully connected layer's matrix. Under the assumption that both fully-connected layers have linear activation functions, the model can be mathematically depicted as a basis projection. Hence, the fully connected layers act as projection schemes to (a) compress input data to a latent space of low dimensionality, and (b) project the LSTM prediction to physical space. 

%An important observation is that the input and output latent spaces the LSTM layer operates in, do not have to be the same. This is particularly advantageous when the input and output flows have significantly different statistics. Hence, this model resembles standard notions of utilizing basis functions to project spatial data to latent dimension. 
\par
The used loss function is a standard mean-square error (MSE) loss 
\begin{gather}\label{eq:MSE_loss}
    \mathcal{L} = \alpha \sum_t \sum_{\phi} \sum_{\theta} \cos \left( 2\pi \frac{\theta}{360} \right) \lVert \hat{\mathbf{X}} -\mathbf{X}^{\text{ref}} \rVert^2,
\end{gather}
where $\alpha$ is a normalization coefficient. Each term in the sum is multiplied by a cosine that is a function of the latitude to showcase that the integration takes place over a sphere. If that term is absent, the model would over-emphasize on learning the corrections at the poles.

\section{Application on E3SM Data}
The numerical exploration of the proposed method begins with some validation results. Specifically, nudged data not used for training are used as input, with the purpose of checking whether the produced output has good accuracy. Training takes places over 1000 epochs and~\cref{eq:MSE_loss} is used as the loss function. Training took place over the time-period 2007-2011 using only Nudged data as input, not CLIM data. Year 2012 was used for validation during training. In~\cref{fig:UV_Zonal_Averages}, results regarding zonally averaged predictions of zonal and meridional velocities $U$ and $V$ respectively, are shown. The left column of subfigure (a) displays the zonally averaged predictions of zonal velocity $U$, for ERA5 reanalysis data, Nudged data and neural network predictions. The right column displays the biases compared to ERA5 predictions for CLIM, Nudged and the neural network predictions. Similarly, the left column of subfigure (b) displays the zonally averaged predictions for meridional velocity $V$, with the same biases displayed. Results are shown for sigma-level $71$, i.e. the one closest to the surface of the earth. This level was then one CLIM predictions differed the most with respect to ERA5 reanalysis data. The results show that the neural network clearly is able to learn a correction for both velocity components. 
\par
Now, the model is tested on unseen free-running and coarse-scale climate models. This dataset (CLIM) is not nudged and thus does not include any ERA5 information. Since the free-running dataset diverges from ERA5 in terms of phases, specific extreme events cannot be studied, as they are absent from CLIM. Hence, long-time statistics are studied now. 
\par
\begin{figure*}
    \centering
    \subfloat[]{{\includegraphics[width=0.40\textwidth]{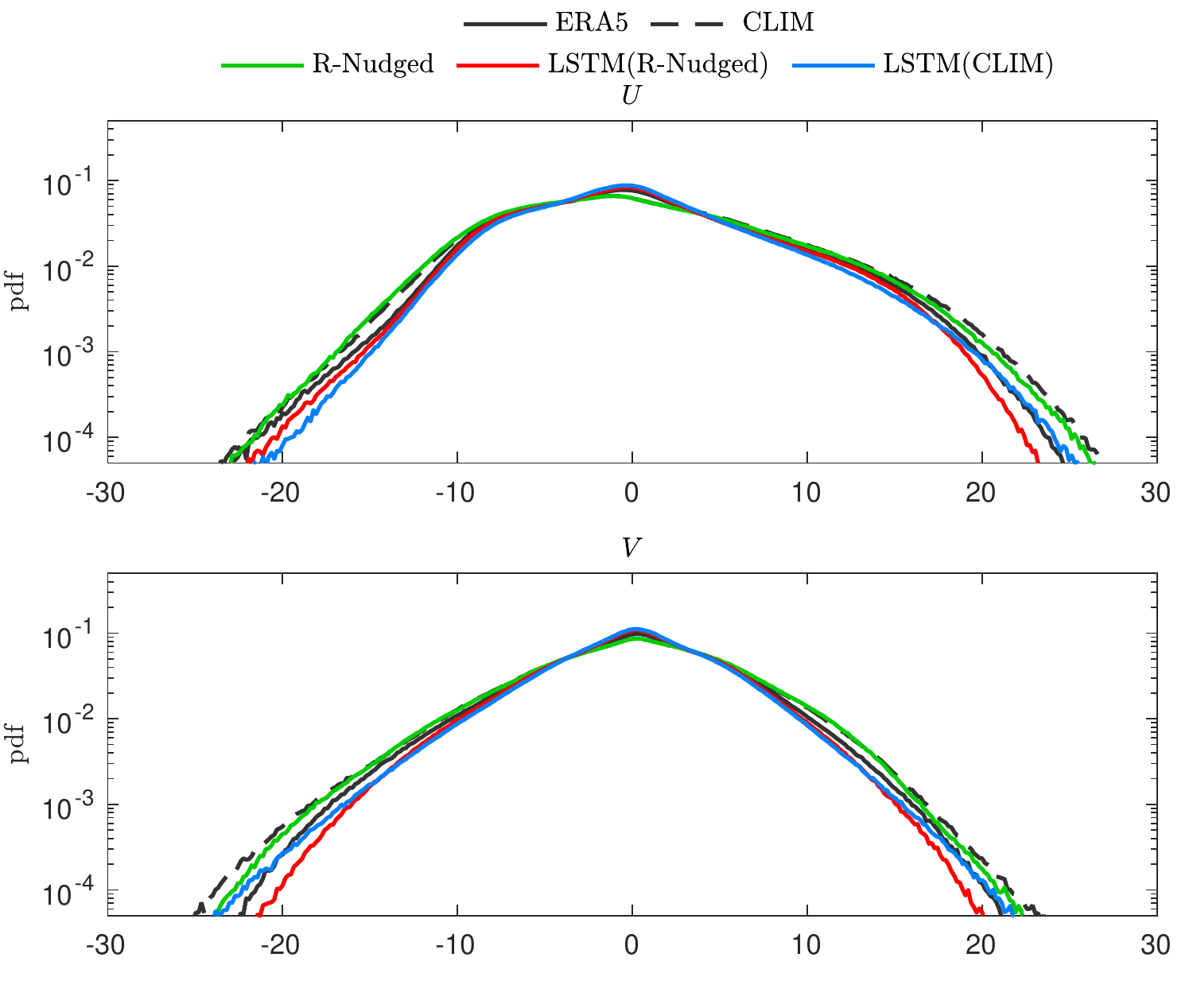} }}%
    %\qquad
    \subfloat[]{{\includegraphics[width=0.40\textwidth]{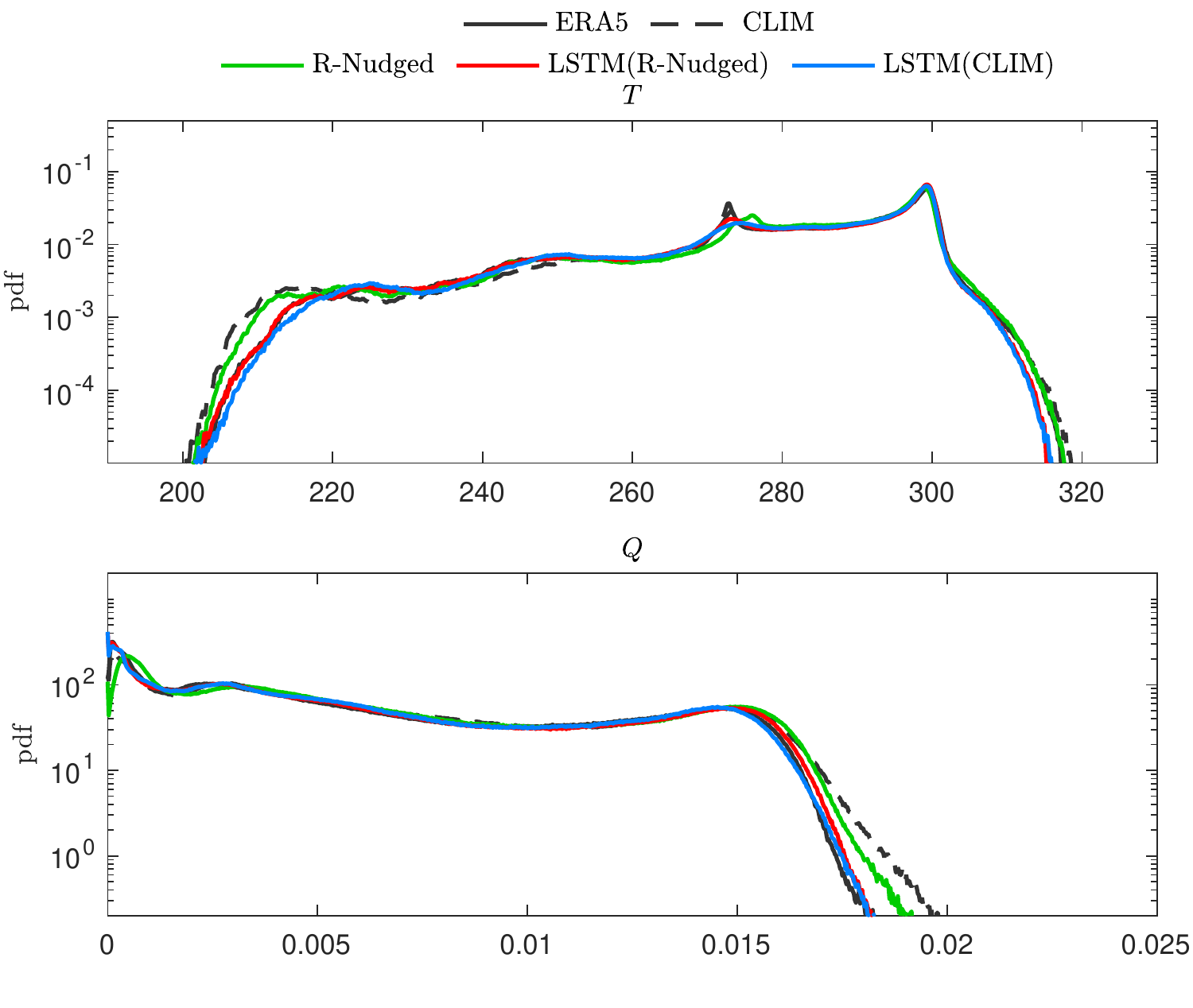} }}%
    \caption{Predicted pdfs for (a) horizontal velocity components $U$ and $V$; (b) temperature $T$ and specific humidity $Q$. Solid black lines correspond to ERA5 data, dashed black lines correspond to CLIM and green lines correspond to Nudged data. Red lines correspond to neural network predictions using Nudged data as input (i.e. training data). Blue lines correspond to neural network predictions using CLIM data as input (i.e. testing data). Results are shown for near-surface data.}%
    \label{fig:Test_PDFs}%
\end{figure*}

In~\cref{fig:Test_PDFs}, the predicted probability density functions (pdf) are shown for the four different predictive variables, $(U,V,T,Q)$. Solid black lines correspond to ERA5 data, dashed black lines correspond to CLIM and green lines correspond to Nudged data. Red lines correspond to neural network predictions using Nudged data as input (i.e. training data). Blue lines correspond to neural network predictions using CLIM data as input (i.e. testing data). Results are shown for sigma-level closest to the earth's surface. Data are averaged over the time-period 2007-2017.
%\begin{figure*}
%    \centering
%    \subfloat[]{{\includegraphics[width=0.40\textwidth]{Figures/PDFUV_Layer70.pdf} }}%
%    %\qquad
%    \subfloat[]{{\includegraphics[width=0.40\textwidth]{Figures/PDFTQ_Layer70.pdf} }}%
%    \caption{Predicted pdfs for (a) horizontal velocity components $U$ and $V$; (b) temperature $T$ and specific humidity $Q$. Solid black lines correspond to ERA5 data, dashed black lines correspond to CLIM and green lines correspond to Nudged data. Red lines correspond to neural network predictions using Nudged data as input (i.e. training data). Blue lines correspond to neural network predictions using CLIM data as input (i.e. testing data). Results are shown for near-surface data.}%
%    \label{fig:Test_PDFs}%
%\end{figure*}
The predictions of the neural network that was trained with Nudged data, significantly improve the prediction of the tails. This can be seen both for the pdfs of the well-predicted quantities $(U,V)$ by CLIM as well as for $(T,Q)$, two quantities whose tails are not well predicted by CLIM. Furthermore, the difference in the predicted pdf when using the training data and free-running data is insignificant, showcasing the ability of the model to generalize beyond training data. This is a result of the pdf of the implemented spectral corrections to the Nudged data, making them display very similar statistics to that of the free running coarse data. This property allows for smoother transfer learning between data sets. 

%\begin{figure*}
%    \centering
%    {\includegraphics[width=1.00\textwidth]{Figures/AR_Density_ARW.pdf} }%
%    \caption{IVT predictions averaged over the period 2007-2017. Mean IVT predictions are shown in the top row for ERA5 data. Biases from ERA5 predictions are shown for, Nudged datasets and CLIM free-running datasets, together with corrected results via our non-intrusive approach. A 1-subregion partition was used for these results.}%
%    \label{fig:AR_IVT_Mean25}%
%\end{figure*}

\begin{figure*}
    \centering
    {\includegraphics[width=1.00\textwidth]{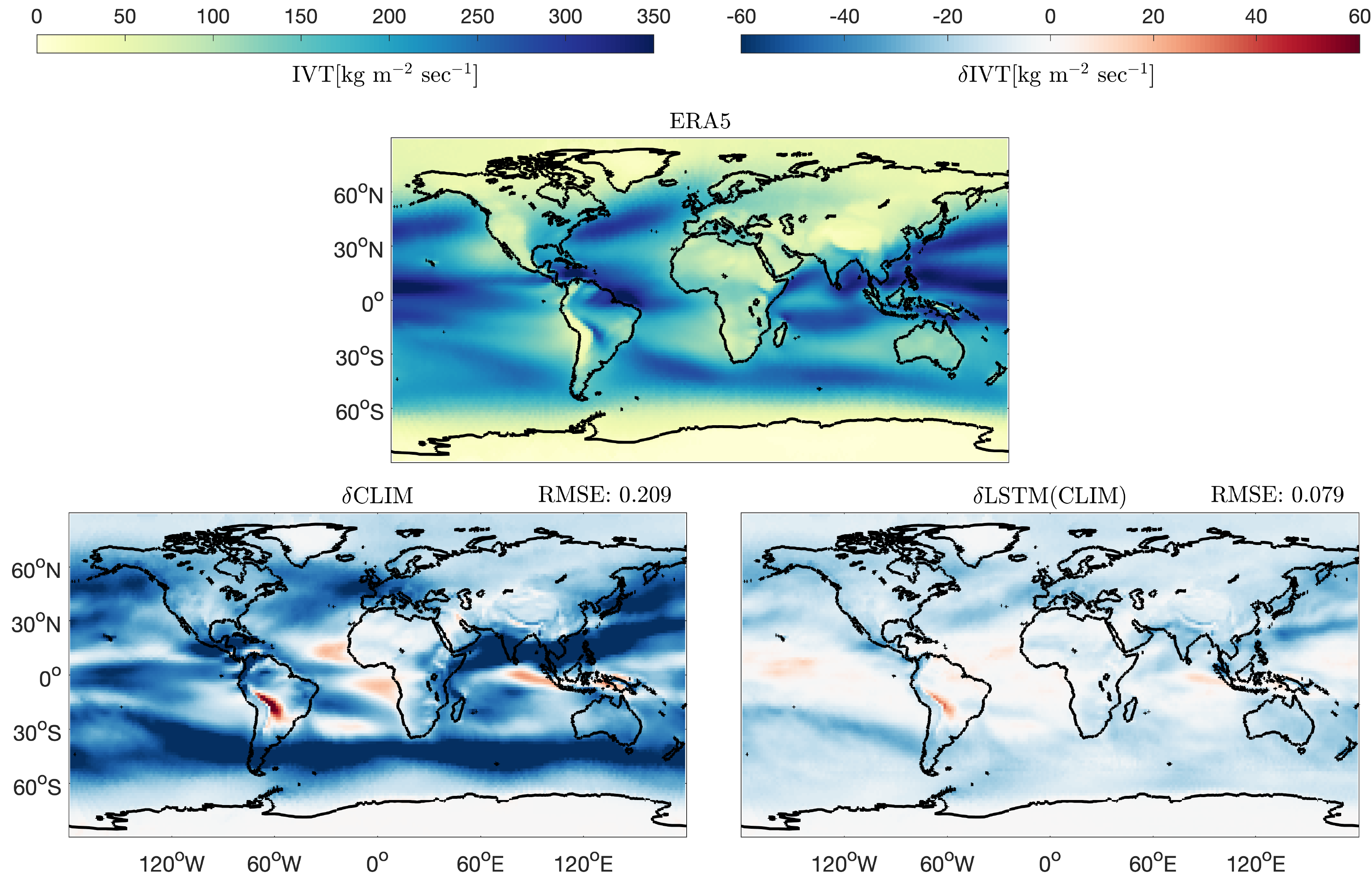} }%
    \caption{IVT predictions averaged over the period 2007-2017. Mean IVT predictions are shown in the top row for ERA5 data. Biases from ERA5 predictions are shown for, the CLIM free-running dataset, together with corrected results via our non-intrusive approach. A 25-subregion partition was used for these results.}%
    \label{fig:AR_IVT_Mean25}%
\end{figure*}

We now move to predict statistics for a derived integral quantity, in particular, mean integrated vapor transport (IVT) over the period 2007-2017. Since IVT is strongly anisotropic, extracting local features, and especially atmospheric rivers, is vital. Therefore the local convolutions derived from each subregion are important for its correction estimation. 25 subregions are used in this numerical test, using different convolutions on each one (i.e. extract local features). In~\cref{fig:AR_IVT_Mean25}, the first row corresponds to ERA5 predictions. The other rows correspond to biases with respect to ERA5. As expected, the neural network exhibits very small biases when Nudged data are used as input. Results for the nudged simulation are not shown for page limit reasons. However, the model significantly decreases biases even when CLIM data are used as input. In addition, it is able to decrease the root mean-squared error (RMSE) below that of the Nudged dataset (which was measured at 0.088), a simulation that exploits ERA5 information at every time-step. 
%This is expected since these data were used for training. However, the model significantly decreases biases even when CLIM data are used as input. In addition, it is able to decrease the root mean-squared error (RMSE) below that of the Nudged dataset, a simulation that exploits ERA5 information at every time-step. 
\par
The final numerical result involves the statistics of tropical cyclones for the time-period 2007-2017. For tracking tropical cyclones, the software package TempestExtreme is used~\cite{ullrich2017tempestextremes,ullrich2021tempestextremes}. The following steps are used to track cyclones: (i) Find local minima of sea-level pressure (SLP). (ii) Eliminate smaller minima within 2 great-circle distance (gcd) degrees. (iii) Check that SLP raises by 200 [Pa] within 8 gcd. (iv) Check that temperature at 400 [mbar] drops by 0.4 [K] within 8 gcd. (v) Check that velocity is higher than 10 [m/sec] for 8 snapshots. (vi) Check that geopotential height is larger than 100 for at least 8 snapshots. (vii) Check that phenomenon lasts at least 54 hours, with a max gap of 24 hours.
\par
\begin{figure*}
    \centering
    {\includegraphics[width=1.00\textwidth]{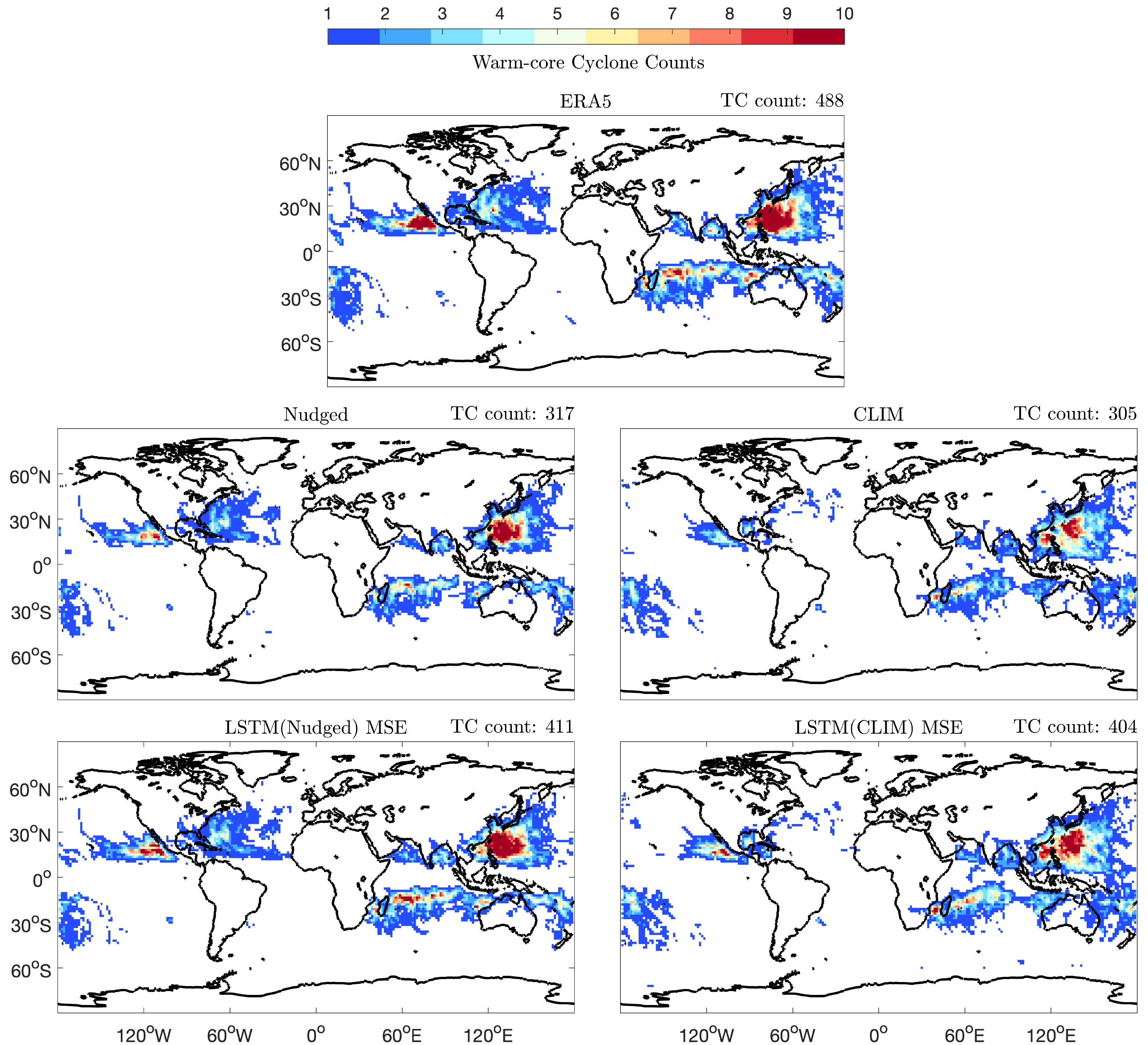} }%
    \caption{Tropical cyclone counts over the period 2007-2017. Results are derived using ERA5 datasets, CLIM free-running datasets and CLIM datasets corrected via our non-intrusive LSTM approach. Cyclones are tracked via the TempestExtremes software.}%
    \label{fig:TC_Cyclone_Counts}%
\end{figure*}

During training, the neural network will track dissipated cyclonic structures in the Nudged simulations, that correspond to tropical cyclones in ERA5. It will then amplify them allowing them to be recognized as tropical cyclones. As a result, if the cyclonic structures present in CLIM are more dissipated than in Nudged, the learned mapping will not be transferable. To overcome this issue, we post process the SLP values of Nudged data. To that end the following conditional means are computed
\begin{gather}
    c(\phi,\theta) = \frac{1}{\text{TC}(\phi,\theta)} \sum_{t=1}^{N_t} \delta SLP (t, \phi, \theta),
\end{gather}
where TC is the tropical cyclone density over the globe. Then, the following spatially dependent coefficient is computed
\begin{gather}
    R(\phi,\theta) = \frac{ c^{\text{c}}(\phi,\theta) }{ c^{\text{c,n}}(\phi,\theta) }.
\end{gather}
Finally, the SLP Nudged data are corrected as follows
\begin{gather}
    SLP^{\text{c,rnsn}} = \begin{cases}
        R(\phi,\theta) SLP^{\text{c,n}}+P_0, & SLP < P_0\\
        SLP^{\text{c,n}}, & SLP \geq P_0
    \end{cases}.
\end{gather}
Results are shown in~\cref{fig:TC_Cyclone_Counts}. In terms of total number of tropical cyclones predicted, the neural network corrections are much better than the ones predicted by the Nudged dataset and CLIM. Nudged data predict 317 tropical cyclones, while CLIM predicts 305. On the other hand, the neural network corrections predict 411 tropical cyclones when using Nudged data and 404 tropical cyclones when using CLIM data. These predictions are much closer to the 488 cyclones predicted from ERA5 data. By focusing on the predictions when using CLIM as input, we notice that barely any new cyclones are predicted in the Atlantic. Looking further into this issue it is our belief that this is a problem of CLIM, where it does not generate enough vorticity over the Atlantic for tropical cyclones to form.

\section{Limitations}
While the proposed methodology was demonstrated to be effective for the prediction of a multitude of climate metrics, some limitations of the current setup should be stated. First, the approach works well under the assumption that the climate is in a statically steady steady, for which a mapping can be learned through the proposed training scheme. Hence, testing the model in simulations where the climate undergoes a transitory phase may hinder its performance, unless similar time intervals are included during training. This is particularly true if the transition is not captured at all by the coarse-scale model. Furthermore, the requirement for reference data (in this case ERA5 reanalysis data), makes the effectiveness of the model unknown under future climate scenarios with drastically different forcings. For such runs to be included in training, high-fidelity simulations would have to be used as reference and nudged towards them. This limitation however is true for online data-driven correction schemes as well since most such models lack concrete error bounds for out-of-sample predictions. Finally, for the application of the scheme to dynamical systems broadly, there is no guarantee that a nudged simulation exists that follows the reference data closely while maintaining the statistics of the coarse simulation. In fact, this is not the case for climate models. However, small discrepancies can be amended via the proposed extra step of the spectral corrections. It should be noted yet, that there is no guarantee the process will work in a case where the deviations from these assumptions are significant. 

\section{Conclusions}
We have formulated and assessed a data-informed hybrid scheme for accurately computing the statistics of climate models. The method employs a nudged solver during training with appropriate spectral corrections to the produced dataset. During testing, the model is assessed on free-running coarse-scale climate models. The approach was applied to realistic atmospheric climate data. Free-running EAMv2 simulations were used as baseline while ERA5 reanalysis data were employed as reference truth. First, the ability of the model to predict global statistics of horizontal velocities $(U,V)$, temperature $T$ and specific humidity $Q$ was assessed. In all cases, the data-informed approach produced results in good agreement with reference ERA5 predictions. Furthermore, the model was tasked with predicting mean IVT over the period 2007-2017, a highly anisotropic quantity. Results are again in good agreement with ERA5, with the corrected dataset having a smaller mean-square error than a nudged dataset. Finally, the model was used to predict statistics of tropical cyclones throughout the globe. The model's ability to correct the number of predicted cyclones over the Pacific was demonstrated.

\subsection*{Acknowledgments} This research has been supported by the DARPA grant HR00112290029 with the program `AI-assisted Climate Tipping-point Modeling' under the program manger Dr. Joshua Elliott. Pacific Northwest National Laboratory is operated for the U.S. Department of Energy by Battelle Memorial Institute under contract. We also thank the two anonymous referees for their constructive comments and feedback.

\bibliography{main_text} 

\end{document}